\documentclass[reprint,
superscriptaddress,
showkeys,
 amsmath,amssymb,
 aps,
]{revtex4-2}
\usepackage{xcolor}
\usepackage{upgreek}
\usepackage{graphicx}
\usepackage{float}
\usepackage{caption} 
\usepackage{soul}

\usepackage{dcolumn}
\usepackage{bm}

\begin{document}

\preprint{APS/123-QED}

\title{Precise \textit{in situ} radius measurement of individual optically trapped microspheres
using negative optical torque exerted by focused vortex beams} 

\author{Kain\~a Diniz}
\email{dinizkaina@gmail.com.br}
\affiliation{Instituto de F\'{\i}sica, Universidade Federal do Rio de Janeiro \\ Caixa Postal 68528,   Rio de Janeiro,  Rio de Janeiro, 21941-972, Brazil}
\affiliation{CENABIO - Centro Nacional de Biologia Estrutural e Bioimagem, Universidade Federal do Rio de Janeiro,
Rio de Janeiro, Rio de Janeiro, 21941-902, Brazil}
\author{Tanja Schoger}
\affiliation{Institut für Physik, Universität Augsburg \\ 86135 Augsburg, Germany}
\author{Guilherme T. Moura}
\author{Arthur L. Fonseca}
\affiliation{Instituto de F\'{\i}sica, Universidade Federal do Rio de Janeiro \\ Caixa Postal 68528,   Rio de Janeiro,  Rio de Janeiro, 21941-972, Brazil}
\affiliation{CENABIO - Centro Nacional de Biologia Estrutural e Bioimagem, Universidade Federal do Rio de Janeiro,
Rio de Janeiro, Rio de Janeiro, 21941-902, Brazil}
\author{Diney S. Ether Jr}
\affiliation{Instituto de F\'{\i}sica, Universidade Federal do Rio de Janeiro \\ Caixa Postal 68528,   Rio de Janeiro,  Rio de Janeiro, 21941-972, Brazil}
\affiliation{CENABIO - Centro Nacional de Biologia Estrutural e Bioimagem, Universidade Federal do Rio de Janeiro,
Rio de Janeiro, Rio de Janeiro, 21941-902, Brazil}

\author{Rafael S. Dutra}
\affiliation{LISComp-IFRJ, Instituto Federal de Educa\c c\~ao, Ci\^encia e Tecnologia, Rua Sebasti\~ao de Lacerda, Paracambi, Rio de Janeiro, 26600-000, Brasil}

\author{Gert-Ludwig Ingold}
\affiliation{Institut für Physik, Universität Augsburg \\ 86135 Augsburg, Germany}

\author{Nathan B. Viana}
\email{nathan@if.ufrj.br }
\affiliation{Instituto de F\'{\i}sica, Universidade Federal do Rio de Janeiro \\ Caixa Postal 68528,   Rio de Janeiro,  Rio de Janeiro, 21941-972, Brazil}
\affiliation{CENABIO - Centro Nacional de Biologia Estrutural e Bioimagem, Universidade Federal do Rio de Janeiro,
Rio de Janeiro, Rio de Janeiro, 21941-902, Brazil}

\author{Paulo A. Maia Neto}
\email{pamn@if.ufrj.br}
\affiliation{Instituto de F\'{\i}sica, Universidade Federal do Rio de Janeiro \\ Caixa Postal 68528,   Rio de Janeiro,  Rio de Janeiro, 21941-972, Brazil}
\affiliation{CENABIO - Centro Nacional de Biologia Estrutural e Bioimagem, Universidade Federal do Rio de Janeiro,
Rio de Janeiro, Rio de Janeiro, 21941-902, Brazil}
\date{\today}

\begin{abstract}
We demonstrate a new method for determining the radius of micron-sized particles 
trapped by a vortex laser beam. The technique is based on measuring 
the rotation experienced by the center of mass of 
a microsphere that is
laterally displaced by a Stokes drag 
force to an off-axis equilibrium position. 
The rotation results from an optical torque pointing along the direction opposite to the vortex beam angular momentum. 
We fit 
the rotation angle data for different Laguerre-Gaussian modes 
taking the radius as a fitting parameter in the Mie-Debye theory of optical tweezers. We also discuss how micron-sized beads can be used as probes for 
optical aberrations introduced by the experimental setup. 
\end{abstract}

\maketitle

\section{Introduction} \label{intro}

The negative optical torque is a remarkable example of a nontrivial exchange of angular momentum between light and matter. 
It arises from the generation of scattered light carrying an excess angular momentum, thus leading
to a recoil torque along the direction opposite to the angular momentum of the incident light. Recent proposals~\cite{Chen2014,Canaguier2015} and experimental
demonstrations~\cite{Hakobyan2014,Magallanes2018,Han2018,diniz2019,Parker2020,Qi2022,Nan2022} cover a wide spectrum of systems. 

\begin{figure}
	\centering 
	\includegraphics[scale=0.4, angle=0]{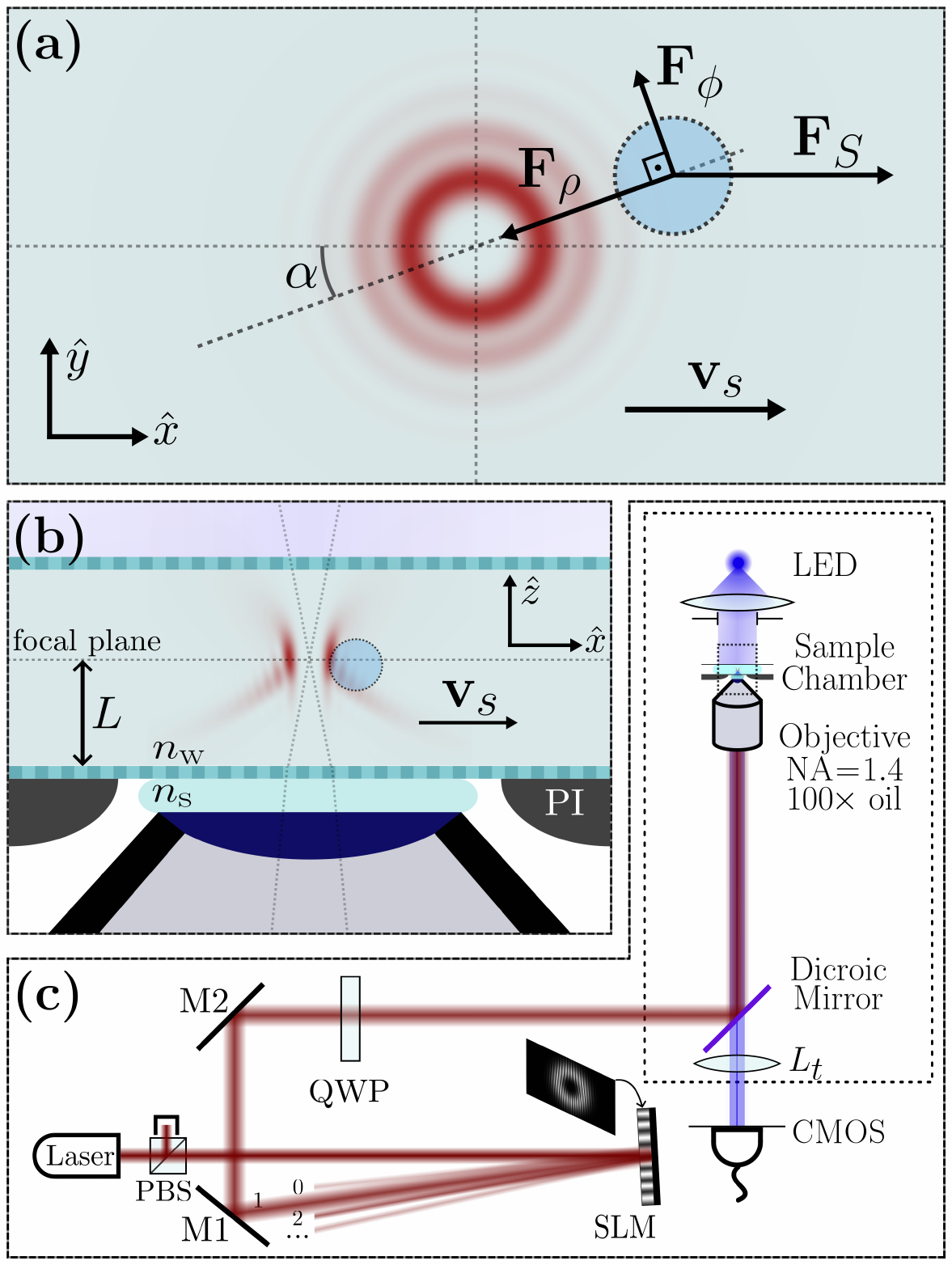}
	\caption{(a) A trapped particle is subject to a constant drag force $\mathbf{F}_{S}$
	 and finds a new off-axis equilibrium position rotated with respect to the direction of this force 
	by an angle $\alpha$. $\mathbf{F}_{\rho}$ and $\mathbf{F}_{\phi}$ denote 
	cylindrical optical force components, and $\mathbf{v}_{S}$ represents the fluid
	velocity. (b) Lateral view of the sample region
 showing the 
    density plot (red) of the electric energy density for the nonparaxial focused beam.
 $L$ denotes the height of the paraxial 
        focal plane with respect to the coverslip. PI represents the piezoelectric nano-positioning system used to move the microscope stage.
        (c) Schematic representation of the experimental setup: The laser beam propagates through a polarizing beam splitter (PBS) towards the spatial light modulator (SLM). The resulting first order of diffraction crosses a quarter-wave plate (QWP)  before entering the microscope (dotted frame). The beam is then focused by an oil-immersion objective into the sample chamber.  
        }
	\label{fig_alpha_setup_data}
\end{figure}

Negative torque experiments usually employ the spin angular momentum associated with circular polarization. In this paper, we use instead the orbital angular momentum of vortex beams which are used to trap a dielectric microsphere 
 in our otherwise typical optical tweezers setup, shown in Fig.~\ref{fig_alpha_setup_data}. Our experimental conditions are such that the laser beam annular focal spot is comparable to or smaller than 
the microsphere radius, thus leading to a stable trap along the beam axis, in contrast to the non-equilibrium steady state of orbital motion measured with smaller beads~\cite{grier2003vortex, Fonseca2023}. We then apply a lateral Stokes force and 
displace the microsphere to a new off-axis equilibrium position as illustrated by Fig.~\ref{fig_alpha_setup_data}(a). The optical torque on the microsphere center-of-mass leads to a rotation of the equilibrium position with respect to the direction of the applied force, which we measure for several values of the vortex beam topological
charge $\ell.$ In most cases, the rotation is opposite
to the handedness defined by the sign of $\ell$.
 
For small values of the Stokes force, the rotation angle is a measure of the 
vorticity of the optical force field near the beam axis. The vorticity is very sensitive to parameters describing the trapped microsphere as well as the focused laser beam. For instance, the chirality of a spherical bead can, in principle, be characterized by measuring the rotation angle~\cite{Ali2020,Ali2020B}. Here, 
we determine the microsphere radius with nanometric precision as we benefit from the enhanced torque exerted by vortex beams. Indeed,
these modes carry an orbital angular momentum $\ell \hbar$ 
per photon that originates from the field's spatial variation~\cite{Allen1992}, 
while the spin angular momentum is limited to $\pm \hbar$ per photon.

The radius of an airborne optically trapped microsphere near a surface was measured by analyzing its Brownian fluctuations~\cite{Burnham2009}.
Precise measurements of airborne microspheres are usually based on Mie resonances, 
taking advantage of the
high quality factor of whispering gallery modes~\cite{Preston2015,McGrory2020}.  However, colloidal particles in a water suspension typically correspond to low refractive index contrasts, thus leading to broader resonances. Dynamic light scattering is a common approach for such systems~\cite{Pecora2000}. Nevertheless, since it relies on measuring intensity fluctuations of light scattered by a sample containing many Brownian particles, this technique does not measure the radius of single individual particles. 
Videomicroscopy is a simple alternative, but inferring 
the size from the light intensity pattern is usually ambiguous, 
particularly for an optically trapped particle. Indeed,  
the microsphere image depends strongly on 
its axial position with respect to the
focal plane~\cite{Gomez2021}, which typically fluctuates by Brownian motion. 

Our approach provides a precise {\it in situ} radius measurement of a specific microsphere which is optically trapped. A comparable precision would be   
obtained by employing electron microscopy. However, in this case 
one would have the inconveniences and risks of isolating the chosen bead, transporting and drying the solution, and coating the sample. 

In addition to measuring properties of the trapped particle, our method also allows us to characterize the astigmatism of the trapping beam. Indeed, there exists a second mechanism  
contributing to the rotation of the equilibrium position.  
When the trapping beam is such as to
break rotational symmetry around the propagation axis, the gradient of the electric energy density is not radial and develops an azimuthal component. 
For example, if the paraxial beam's polarization is linear, 
the nonparaxial focal spot will be elongated along the polarization axis~\cite{RichardsWolf}, resulting in an azimuthal component of the optical force~\cite{diniz2019}.
Optical aberrations that define a preferential direction, such as astigmatism, can also be 
responsible for producing a non-symmetric focal spot~\cite{dutra2012}. 
For that reason, the comparison with the experimental data for circularly polarized focused vortex beams allows us to measure the astigmatism parameters of our setup. In order to
implement such comparison, we develop a theoretical model for the optical force exerted by an astigmatic focused vortex beam.

The paper is organized as follows: Sec.~\ref{sec:exp_procedure} presents the experimental setup and procedure, as well as how the rotation angle is extracted from the raw data. Sec.~\ref{sec:mdsaplus} presents the theoretical model developed to describe the experiments. 
The results are discussed in Sec.~\ref{sec:results_discussion} and concluding remarks are presented in Sec.~\ref{sec:conclusions}. 
Technical details are presented in two appendices. 

\section{Experimental setup and procedure}\label{sec:exp_procedure}

Our method for radius measurements goes as follows. We apply a 
constant Stokes drag force $\mathbf{F}_S$ on a sphere trapped by optical tweezers with a given
Laguerre-Gaussian mode LG$_{p\ell},$ with radial order $p=0$ and topological charge $\ell,$
at the objective entrance port.
The particle is displaced until it finds an off-axis 
equilibrium position rotated by an angle $\alpha$ with respect to the direction of the drag force,
as depicted in Figs.~\ref{fig_alpha_setup_data}(a) and (b). Then, we
vary the topological charge, measuring rotation angles as a discrete
function of $\ell$. Finally, we fit this data to the theoretical model presented in the next section, leaving the radius as a free parameter and, in some cases, also the astigmatism parameters and the paraxial focal height.

The experimental setup for the measurements of $\alpha$ is illustrated in 
Fig.~\ref{fig_alpha_setup_data}(c). A $\rm TEM_{00}$ laser beam (IPG photonics, model YLR-5-1064LP) with vacuum wavelength $\lambda_0 = 1064\,\mathrm{nm}$ illuminates a spatial light modulator (SLM). This device enables the modulation of both the field's amplitude and phase \cite{rosales2017shape}, which allows us to convert the incident beam into chosen $\text{LG}_{0 \ell}$ modes with appropriate waist $w_{0}$ via a complex phase modulation. Notice that we choose a different beam waist for each mode so as to ensure an appropriate filling at the objective's entrance. The criteria for choosing the waists and the method to measure them are described in Appendix~\ref{appendix:filling}, while the  waist values are given in Table~\ref{tab:beam_waist}. Before entering an inverted microscope, the beam is left-polarized by a quarter-wave plate. 
It is then focused by a $100\times$, oil-immersion objective with numerical aperture $\rm NA = 1.4$ and back aperture
radius $R_{\rm obj} = 2.8 \, \rm mm$
through the glass-water interface
into an aqueous dispersion of polystyrene beads (Polysciences), where the optical trapping occurs (see Fig.~\ref{fig_alpha_setup_data}(b)). 
This solution is illuminated by a 470\,nm blue LED whose scattered light, after being collected by the objective, is recollimated by a tube lens ($L_t$) to form an image on a CMOS camera 
(Hamamatsu Orca-Flash 2.8 C11440-10C).

\begin{figure}
	\centering
	\includegraphics[width=0.45\textwidth]{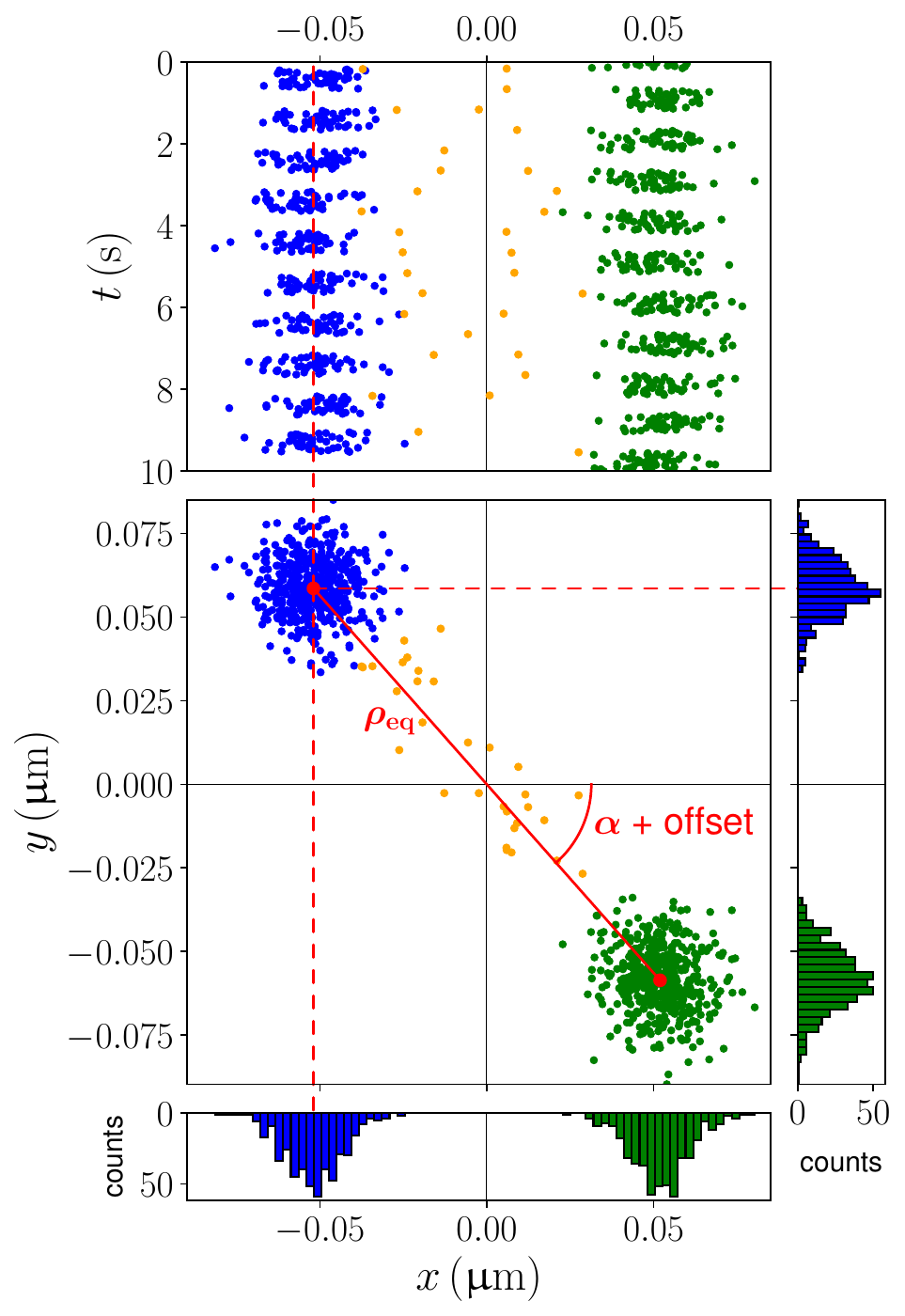}
	\caption{The $x$ and $y$ coordinates 
of the bead are measured over a time of
    $T=10\,\mathrm{s}$. The upper panel shows a typical distribution of the 
    measured $x$ coordinates. Green and blue clusters correspond to opposite 
    velocities of the microscope stage. The central panel shows 
    the measured positions on the $xy$ plane. The histograms for the 
    $x$ and $y$ coordinates are fitted by Gaussian functions in order to determine the 
    equilibrium positions under the Stokes force (red points). 
	The rotation angle $\alpha$ is then obtained from the slope of the red line 
	connecting the two equilibrium positions. }
	\label{fig:exp_data}
\end{figure}

After trapping a microsphere with a focused Gaussian beam ($\ell=0$), we first lower the objective until the microsphere touches the bottom of the sample chamber. Then, we move up the objective by a height of $d=(2 \pm 1)\, \upmu \mathrm{m}$
 so as to define a new trapping position a few microns above the coverslip. To apply a constant drag force, we use a piezoelectric nano-positioning stage (Digital Piezo Controller E-710, Physik Instrumente)
to move the microscope stage with the velocity
${\bf v}_S=\pm v_S\,\mathbf{\hat x},$ $v_S=20\,\upmu{\rm m/s},$
for 0.5\,s, alternating 
between the positive and negative $x$ directions. At each cycle, the Stokes drag force displaces the
particle from its on-axis equilibrium position. In a given run, the bead's movement 
is recorded for 10\,s. The center-of-mass position on the $xy$-plane at each frame is later 
determined by video analysis with Fiji~\cite{FijiRef}, as depicted by way of example  
for the $x$ coordinate in the upper panel of Fig.~\protect\ref{fig:exp_data}.

The camera itself is rotated by an angle close to $45^\circ$ to ensure that the displacements along 
the $x$ and $y$  directions are roughly of the same magnitude. 
The central panel in Fig.~\ref{fig:exp_data} depicts all particle positions obtained
from the measured $x$ coordinates shown in the upper panel as well as from the corresponding $y$ coordinates
(not shown) for a typical run. 
All recorded positions are categorized into three separate
clusters by means of a Scikit-learn routine \cite{scikit-learn} and marked as blue, green, and
orange points in Fig.~\ref{fig:exp_data}. 
In this way, the orange points that lie in between
the two main clusters are identified and no longer considered for the analysis.

The marginal distributions for the $x$ and $y$ directions can be 
fairly described by Gaussian
distributions so that we fit the two main clusters to two-dimensional
Gaussians in order to determine the coordinates of the two equilibrium positions.  We then obtain the angle depicted in Fig.~\ref{fig:exp_data} 
between the line connecting 
the two equilibrium positions and the $x$ direction. Its error is computed by propagating the error of the coordinates. Finally, to determine the rotation angle $\alpha$
depicted in Fig.~\ref{fig_alpha_setup_data}(a), we subtract the camera offset angle, which is determined by recording the position of a fixed reference spot in the coverslip as the microscope stage is driven.

Apart from the rotation angle, we also extract the radial displacement $\rho_{\rm eq}$ of the 
sphere from the focal point. By assuming that the movement of the bead is 
symmetric, we can compute the displacement from the distance
between the two equilibrium positions. 

For each beam mode $\ell$ the measurement was repeated several times.

\section{MDSA+ theory of the optical force by a focused vortex beam}\label{sec:mdsaplus}

The optical force acting on an illuminated particle is generally computed by integrating the time-averaged Maxwell stress tensor over a surface which encloses the particle. 
The stress tensor includes the incident and scattered field contributions. Within the Mie-Debye theory~\cite{Neto2000,mazolli2003}, the former is described by a Debye-type 
nonparaxial model~\cite{RichardsWolf} for a tightly focused beam, while the latter follows from standard Mie scattering by a spherical particle. An extension to include the spherical aberration introduced by focusing through the 
glass-water planar interface led to the MDSA (Mie-Debye with spherical aberration) 
theory~\cite{viana2007,Dutra2007}.
In \cite{dutra2014}, the model was further 
extended (MDSA+ theory) to allow for the presence of any primary aberration on the paraxial Gaussian beam before focusing. 

To mimic the experimental setup discussed in Sec.~\ref{sec:exp_procedure}, here 
we consider
that the laser beam at the objective entrance port is a 
Laguerre-Gaussian mode LG$_{0\ell}.$ 
Expressions for the aplanatic focusing of a Laguerre-Gaussian beam were developed in \cite{monteiro2009}. Building on these 
earlier results, we find the following angular spectrum decomposition for the electric field in the aqueous solution
obtained from focusing a 
 circularly polarized LG$_{0\ell}$ beam at the objective entrance port:
\begin{equation}
\begin{aligned}
\mathbf{E}^{(\sigma)}_\ell(\mathbf{r}) &= -\frac{ikf E_0 e^{-ikf}}{2\pi} \left(\sqrt{2}\gamma \right)^{|\ell|}
	\int_0^{2\pi} d\varphi \, e^{i\ell \varphi} \\
	& \quad \times
	\int_0^{\theta_m} d\theta\, \sin\theta \sqrt{\cos\theta} 
	\sin^{|\ell|}(\theta) 	
	e^{-\gamma^2 \sin^2\theta} \\
	& \hspace{4em} \times
	T(\theta)
	e^{i (\Psi_\text{g-w} +  \Psi_\text{ast})}
	e^{i\mathbf{k}_\text{w} \cdot \mathbf{r}}
	\hat{\epsilon}_\sigma (\theta_\text{w}, \varphi_\text{w})\,.
\end{aligned}
\label{eq:E_inc}
\end{equation}
The spherical components $(k, \theta, \varphi)$ describe the 
wave vectors in the glass slide. 
The parameter $\gamma = f/w_{0}$ defines the ratio of the objective 
focal length $f$
and the beam waist $w_{0}$ at the objective entrance port. 
The polar angle $\theta_\text{w}$ in the host medium is defined through 
Snell's law: $\sin\theta_\text{w} = \sin\theta/N_\text{s}$ 
where $N_\text{s}=n_\text{w}/n_\text{s}$ is the 
relative refractive index between the fluid and the glass slide.  
The integration is performed up to a maximum angle $\theta_\text{m}$ given by
$\sin\theta_\text{m} = \text{min}\left(N_\text{s}, \sin\theta_0\right)$ with
$\sin\theta_0 = \mathrm{NA}/n_\text{w}$, where NA denotes the numerical aperture
of the objective. 
The wave vector $\mathbf{k}_\text{w}$ in the sample is characterized by its modulus 
$k_\text{w} = N_\text{s} k$ and the spherical angles $(\theta_\text{w}, \varphi_\text{w})$ where $\varphi_\text{w} = \varphi$. 
The unit vector $\hat{\epsilon}_\sigma(\theta_\text{w}, \varphi_\text{w}) = e^{i\sigma \varphi_\text{w}}(\hat{\theta}_\text{w}
 + i\sigma \hat{\varphi}_\text{w})/\sqrt{2}$ accounts for a right- ($\sigma=-1$) or left-handed ($\sigma = 1$) circularly 
polarized Fourier component along the propagation direction $(\theta_\text{w}, \varphi_\text{w}).$
It is obtained by rotating the unit vector
$(\hat{x}+i\sigma\hat{y})/\sqrt{2}$ at the entrance of the objective by the Euler angles 
$(\varphi_\text{w}, \theta_\text{w}, -\varphi_\text{w}).$

The astigmatism (ast) introduced by the optical elements of the experimental setup is accounted for by the Zernike phase \cite{zernike1934, born1959}
\begin{equation}
\label{Zernike-astigmatism}
\Psi_\text{ast} = 2\pi A_\text{ast} 
\left(\frac{\sin\theta}{\sin\theta_0}\right)^2 \cos\left[2(\varphi - \phi_\text{ast})\right]
\end{equation}
given in terms of the  amplitude $A_\text{ast}$ and of the angle $\phi_\text{ast}$ defining the 
astigmatism axis and thus breaking rotational symmetry.

We also take into account the spherical aberration introduced by refraction at
the interface between the coverslip and the aqueous suspension. 
Neglecting the dependence of the Fresnel refraction coefficients on the polarization, the transmission amplitude is \cite{viana2007}
\begin{equation}
T(\theta) = \frac{2\cos\theta}{\cos\theta + N_\text{s} \cos\theta_\text{w}}
\end{equation}
whereas the spherical aberration phase reads \cite{Torok95}
\begin{equation}
\label{spherical_aberration}
\Psi_\text{g-w} = k L\left(N_\text{s} \cos\theta_\text{w} 
		- \frac{\cos\theta}{N_\text{s}}\right)
\end{equation}
where $L$ is the distance between the glass-water (g-w) interface and the paraxial focus
as illustrated in Fig.~\ref{fig_alpha_setup_data}(b).

The incident field \eqref{eq:E_inc} is inserted into the Maxwell stress tensor together with the scattered field, which is obtained by solving the scattering problem 
using Mie theory for spherical particles combined with the appropriate Wigner rotation matrix elements. 
The integration over the stress tensor can be worked out analytically. 
Explicit expressions for the cylindrical force components ($F_\rho, F_\phi, F_z$) as functions of the position of the sphere center $\mathbf{R}(\rho,\phi,z)$ with respect to the focus can be found in Appendix~\ref{sec:force_components}.

\section{Results and discussion}\label{sec:results_discussion}

\subsection{Numerical analysis}

The sphere radius $R$ as well as the parameters for the optical aberrations
can be obtained by fitting the experimental data
with the rotation angles obtained from the MDSA+ theory. Two optical aberrations are considered in our model. The first one is the spherical aberration produced 
at the glass-water interface, which is 
characterized by the phase (\ref{spherical_aberration}) proportional to the distance $L$ of the
paraxial focus from the coverslip. The second is astigmatism, defined by 
(\ref{Zernike-astigmatism}) in terms of 
the amplitude $A_\text{ast}$ and the angle of the principal meridian $\phi_\text{ast}$. The distance $L$ can, in principle, be determined by emulating the experimental procedure described in Sec.~\ref{sec:exp_procedure}~\cite{viana2007}.
Considering a Gaussian mode at the entrance 
of the objective, we first compute the 
initial focal height $L_0$ for which the 
on-axis equilibrium position 
is such that the sphere is 
touching the coverslip. The equilibrium position is obtained
from the requirement of a vanishing axial force component
\begin{equation}
F_z^{(\ell =0)}\left(\rho=0, \phi=0, z = R-L_0, L=L_0\right) =0 \,.
\end{equation}
Afterwards, the focal plane is displaced upwards by $d$ leading to the final height
$L = L_0 + N_s d$.
The rotation angle for each mode $\ell$ can then be found by solving the following system of equations for 
the equilibrium position $\mathbf{R}_\text{eq} = \mathbf{R}_\text{eq}(\rho_\text{eq}, \alpha, z_\text{eq})$ under the applied Stokes drag force:
\begin{equation}
\begin{aligned}
F_z^{(\ell)}(\mathbf{R}_\text{eq}) &=0 
\\
\arctan\left(\frac{F_\phi^{(\ell)}(\mathbf{R}_\text{eq})}
{F_\rho^{(\ell)}(\mathbf{R}_\text{eq})}\right) &= \alpha
\end{aligned}
\label{eq:eq_system}
\end{equation}
with the force expressions given in Appendix~\ref{sec:force_components}. 
The rotation angle $\alpha$ corresponds to the azimuth angle $\phi$ of the bead at its equilibrium position. 
The radial component $\rho_\text{eq}$ is extracted from the experimental data as described in the previous
section. 
We found that the rotation angle only weakly depends on $\rho_\text{eq}$, meaning that the changes in the rotation angle with the radial distance are negligible. For each $\text{LG}_{0\ell}$ mode, we thus average $\rho_\text{eq}$ over all measurement runs for a given bead and use the resulting average value as input for the radial equilibrium coordinate.

Note that without astigmatism ($A_\text{ast} = 0$)
 the force components do not explicitly depend on the azimuthal angle $\phi$. 
Hence, the set of equations \eqref{eq:eq_system} no longer needs to be solved simultaneously. 
Instead, after determining the axial equilibrium position, we can directly calculate the rotation angle. 
The computation time using only the MDSA theory is thus significantly reduced compared to the full calculation within the MDSA+ theory. 

The fitting is done by minimizing a weighted sum of squared errors
between the theoretical and experimental results for the rotation angle
\begin{equation}\label{eq:chi2}
\chi^2 =  \frac{1}{N}\sum_\ell \frac{1}{M_\ell} \sum_{r} \left(\frac{\alpha(\ell) - 
\alpha_\text{exp}(\ell, r)}{\Delta \alpha_\text{exp}(\ell, r)}\right)^2
\end{equation}
where $\alpha_\text{exp}(\ell, r) \pm \Delta \alpha_\text{exp}(\ell, r)$ is the experimentally obtained rotation angle and its error for the beam mode $\ell$ and run $r$,  $N$ is the number of modes used for fitting and $M_\ell$ is the number of 
measurement rounds performed for each beam mode. The maximum number of runs $M_\ell$ was ten, while in a few cases, due to experimental reasons, only data from one run could be used.

The force calculation was implemented in Python using the scientific libraries 
NumPy \cite{harris2020} and SciPy \cite{SciPy2020}.
For the minimization of $\chi^2$ we used the iminuit package \cite{iminuit}. 

Relevant parameters for the simulations not yet mentioned include the refractive index of the immersion oil 
$n_\text{s} = 1.518$ and the refractive index of the
polystyrene beads $n = 1.5694$, which we obtained by linearly interpolating the data given in \cite{zhang2020} to our wavelength of $\lambda_0 = 1064\,$nm.
The laser power, below $120\,{\rm mW}$ at the objective entrance
port, is too low to heat the sample. Therefore, we assume it to be in thermal equilibrium with its environment and use the refractive index for water $n_\text{w}=1.3246$ at $T=19\, ^\circ$C
in our calculations. For comparison, we also performed the calculation for $T=24\, ^\circ$C with
$n_\text{w} = 1.3242$ and found no significant change in the fitted parameters. These refractive indices for 
$\lambda_0$ were obtained from a linear interpolation of the 
data given in \cite{daimon2007}.

\subsection{Beads of nominal radius 1.5 microns}

\begin{figure}
	\centering 
	\includegraphics[width=0.45\textwidth]{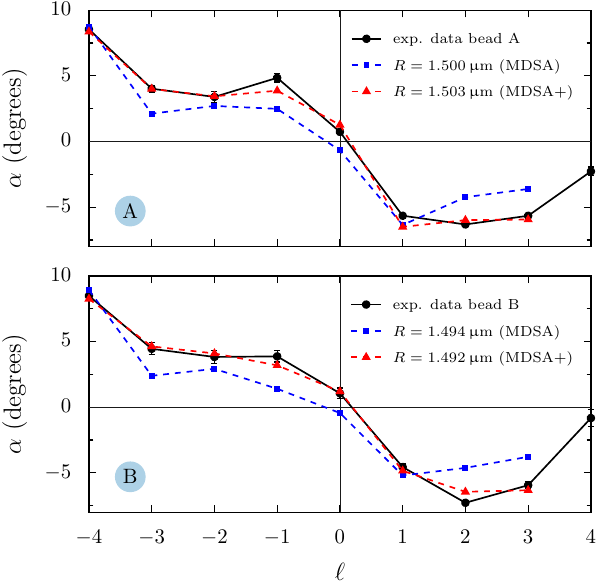}	
	\caption{Microsphere rotation angle $\alpha$ versus beam mode $\ell$ 
		for beads A and B. The circles depict the experimental results and 
            the squares and triangles are the result of fitting with MDSA and MDSA+ theories, respectively, with the latter including the effect of astigmatism. In all plots, we have added lines connecting the data points as a guide to the eye.
		} 
	\label{fig:beadAB}
\end{figure}

We begin by discussing the results for two beads whose nominal radius is $(1.50 \pm 0.04) \,\upmu \mathrm{m}$.
In the following, we refer to them as beads A and B. 
Fig.~\ref{fig:beadAB} shows the experimental results for the rotation angles of each 
bead for beam modes ranging from $\ell=-4$ to $\ell=4$. 
The error bars are the standard deviations of $\alpha$ computed from the multiple rounds 
of measurement. 

For each sphere, we performed a fit, 
where we excluded the $\ell=4$ case for which we found no reasonable value for $\alpha$.
We believe that the reason for this is that $\ell = 4$ is close to 
the bistable trapping regime demonstrated in Ref.~\cite{Fonseca2023}.
 Although the bead does 
have stability around the axis, it also probes a region
of positive torque near the annular focal spot, due to its thermal fluctuations. This causes the experimental value of $\alpha$ to be much less negative than the value of about $-15.7^\circ$ predicted by theory.

First, we fitted the experimental curve by just using the MDSA theory 
with $L$ determined as described above.
The fitted radii can be found in Tab.~\ref{tab:fitAandB_mdsa} and $\alpha$ as a function of $\ell$ is depicted in Fig.~\ref{fig:beadAB}. 
Overall, the fitted data points show good qualitative agreement with the experimental result, 
except for $\ell=0$. 
In this case, theory predicts a negative rotation angle, while the experimentally 
observed one is positive. 
We will discuss the reason for this discrepancy below. 

\begin{table}
\centering
\setlength{\tabcolsep}{5.5pt} 
\renewcommand{\arraystretch}{1.2}
\caption{Optimal radii for beads of nominal radius $(1.50 \pm 0.04) \, \upmu \mathrm{m}$ from a fit within the MDSA theory (zero astigmatism).}
\begin{tabular}{ccc}
\hline\hline
bead & $R(\upmu \mathrm{m})$ & $\chi^2 $  \\
 \hline 
A & 1.500 $\pm$ 0.004 & 21.7\\
B & 1.494 $\pm$ 0.003 & 26.4 \\
\hline\hline
\end{tabular}
\label{tab:fitAandB_mdsa}
\end{table}

Next, we used the MDSA+ theory for fitting with $A_\text{ast}$, $\phi_\text{ast}$ 
and $L$ as additional parameters. 
The parameters obtained from the fit can be found in Tab.~\ref{tab:fitAandB}.
The fitted data points depicted in Fig.~\ref{fig:beadAB} show almost perfect agreement with 
the experimental results. Notice that the $\chi^2$ value does not indicate
overfitting. In addition, in spite of the large number of parameters,
the two independent fits for beads A and B found astigmatism parameters $A_\text{ast}$ and $\phi_\text{ast}$
 that agree within error bars. 
 This is to be expected since
the astigmatism is a characteristic of the experimental setup, thus remaining the same for measurements with both beads. On the other hand, the difference of about 1\,$\upmu \mathrm{m}$ 
between the values of $L$ for the two beads is still consistent with the large error of adjusting the height of the objective as mentioned in Sec.~\ref{sec:exp_procedure}. Following this reasoning, we also 
performed a joint fit for beads A and B with shared
parameters for the astigmatism. The results can be found in Tab.~\ref{tab:jointfitAB}. 

As for the fitted radii, notice that they are close to the ones found by fitting the data to the MDSA theory. This shows that if one wishes solely to characterize a bead's radius, the method can be performed with this simpler version of the theory, drastically reducing the necessary computation time. Furthermore, we have performed the MDSA-fit with each individual round and found values for the radius compatible with the ones in Tab.~\ref{tab:fitAandB_mdsa}, suggesting that the method could be greatly simplified.


\begin{table*}
\centering
\setlength{\tabcolsep}{10pt} 
\renewcommand{\arraystretch}{1.2}
\caption{Fit parameters for beads of nominal radius $(1.50 \pm 0.04) \, \upmu \mathrm{m}$.}
\begin{tabular}{cccccc}
\hline \hline
bead &  $R (\upmu \mathrm{m})$ &  $L (\upmu \mathrm{m})$ & $A_\text{ast}$ 
	& $\phi_\text{ast} \text{(rad)}$  &  $\chi^2$
\\ \hline 
A &  1.503 $\pm$ 0.007  & 5.8 $\pm$ 0.6  & 0.245  $\pm$ 0.026  
	& 0.37  $\pm$ 0.14  & 2.2
 \\
B &  1.492 $\pm$ 0.005  & 4.7 $\pm$ 0.6 & 0.262 $\pm$ 0.024
	& 0.44  $\pm$ 0.12 & 2.3	
	\\
	\hline\hline
\end{tabular}
\label{tab:fitAandB}
\end{table*}


It should be noted that only the presence of higher-order modes allows the distinction between spheres A and B. 
The measured rotation angles for a Gaussian mode are 
$\alpha_\mathrm{A}(\ell=0) = (0.7 \pm 0.2)^\circ$ 
and $\alpha_\mathrm{B}(\ell=0) = (1.1 \pm 0.4)^\circ$
respectively for beads A and B. They are practically indistinguishable 
within error bars. 
However, if we now look at the case $\ell = 2$, the corresponding results
are $\alpha_\mathrm{A}(\ell=2) = (-6.3 \pm 0.1)^\circ$ 
and $\alpha_\mathrm{B}(\ell=2) = (-7.3 \pm 0.1)^\circ$. 
Note that in that case not only are the values further apart, but they are 
distinguishable beyond error bars. 

Finally, we emphasize that for all modes except $\ell = 0$ the angular momentum gained by the particle is opposite to that carried by the paraxial beam. This exception happens because for this mode the total angular momentum per photon is relatively small, and the symmetry-breaking effect of astigmatism dominates
over the angular momentum exchange. This is corroborated by $\ell = -1$, where the total angular momentum per photon is zero, and the torque, caused exclusively by the spot-asymmetry, is positive. Therefore, the negative optical torque reported in \cite{diniz2019} extends to $\rm LG_{0 \ell}$ modes and seems to be a characteristic of near-focus interactions, rather than a particularity of that setup.
 

\begin{table*}
\centering
\setlength{\tabcolsep}{5.5pt} 
\renewcommand{\arraystretch}{1.5}
\caption{Parameters from the joint fit for two beads with nominal radius $(1.50 \pm 0.04) \, \upmu \mathrm{m}$
using shared values for the
astigmatism. }
\begin{tabular}{ccccccc}
\hline\hline
  $R_A (\upmu \mathrm{m})$ & $R_B (\upmu \mathrm{m})$ 	
	& $L_A (\upmu \mathrm{m})$ & $L_B (\upmu \mathrm{m})$ 
	& $A_\text{ast}$ & $\phi_\text{ast} \text{(rad)}$  
	&  $\chi^2_{A,B}$
\\ \hline 
 1.502 $\pm$ 0.010 & 1.492 $\pm$ 0.007  
	& 5.56 $\pm$ 0.83  & 4.85 $\pm$ 0.78
	& 0.254 $\pm$ 0.026  & 0.41 $\pm$ 0.14
	& 2.4, 2.4  
	\\
\hline \hline
\end{tabular}
\label{tab:jointfitAB}
\end{table*}


\subsection{Beads of nominal radius 2.260 microns}

\begin{figure}
\centering
\includegraphics[width=0.45\textwidth]{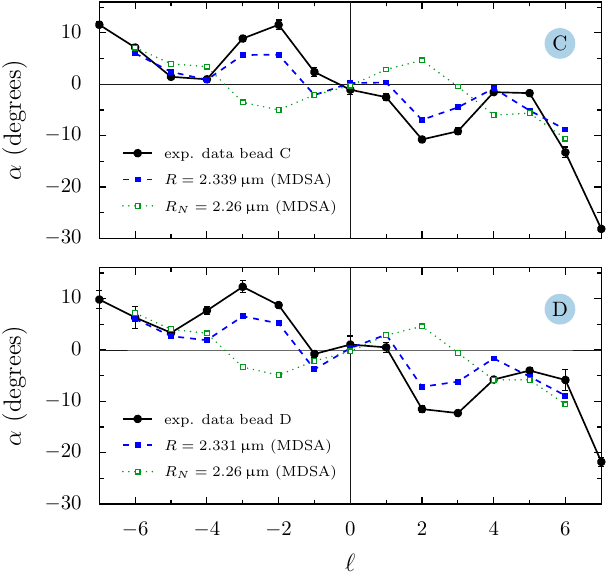}
\caption{Rotation angles for beads C and D as functions of the beam mode $\ell$.
The circles depict the experimental results.
The fitted data from the MDSA theory is shown in blue. 
The values at $\ell = \pm 7$ were excluded from the fit. As in Fig. \ref{fig:beadAB}, the lines connecting the data points serve as a visual aid.  
}
\label{fig:beadCD}
\end{figure}

We now discuss the results for two beads whose radii have a nominal value of 
$R_\mathrm{N} =  (2.260 \pm 0.075)\, \upmu \mathrm{m}$ and are referred to 
as beads C and D in the following. Due to their larger size, these beads were stably trapped on the optical axis for a broader range of topological charges~\cite{Fonseca2023} when compared to the smaller beads discussed above. The experiments were performed with $\ell$ ranging from $-7$ to $7$ 
yielding the data shown in Fig.~\ref{fig:beadCD}. 
Beads in the geometrical optics regime $R \gg \lambda$ are less sensitive to optical aberrations because the asymmetries in the field distribution are averaged out over the sphere.
We thus neglect the astigmatism and use fixed values for 
the distance $L$ as determined from the known displacement $d$ of the objective when fitting the experimental data.

The experimental data for $\ell$ ranging from $-6$ to $6$ were fitted. In the cases $\ell= \pm 7$, 
we could not find equilibrium positions for all radii in the fitting interval ranging from 2.11 to 2.41\,$\upmu$m. 
We believe that the explanation lies again in the size of the annular focal spot, 
which is too large to allow for stable trapping on the beam axis for the highest values of $|\ell|$. 
The radii obtained from the fit are given in Tab.~\ref{tab:beadCD} and 
the fitted data for $\alpha$ are depicted in Fig.~\ref{fig:beadCD}.
For comparison, we also performed the fit using the MDSA+ theory with the astigmatism parameters 
obtained from the joint fit of beads A and B. 
The resulting radii lay within the error bars of the radii obtained within the MDSA theory
and the quality of the fit shows no significant improvement. 
Due to the high relative error of measuring the displacement $d$, we also performed the fit with
$d=3\,\mu \mathrm{m}$ which again leads to the same optimal radii. 
Our findings thus confirm that both astigmatism and spherical aberration are less critical for larger spheres, which implies that here the fitting process is even more robust than for smaller spheres.

The sensitivity of our method with respect to changes in the radii can be inferred from Fig.~\ref{fig:beadCD}, where the rotation angles are also depicted for the nominal radius $R_{\mathrm{N}}$. 
The agreement with the experimental data becomes much worse in this case and the qualitative aspects of the curve change, with even an opposite sign for the rotation angles for modes with topological charges in the range $-3 \leq \ell \leq 2$.

\begin{table}
\centering
\setlength{\tabcolsep}{5.5pt} 
\renewcommand{\arraystretch}{1.2}
\caption{Optimal radii for beads of nominal radius $(2.260 \pm 0.075) \, \upmu \mathrm{m}$.}
\begin{tabular}{cccc}
\hline\hline
bead &  $R (\upmu \mathrm{m})$ &  $\chi^2$
\\ \hline 
C &  2.339 $\pm$ 0.003  &  40.1
 \\ 
D &  2.331 $\pm$ 0.003 & 67.7
 \\
 \hline \hline
\end{tabular}
\label{tab:beadCD}
\end{table}

\begin{figure}
\centering
\includegraphics[width=0.45\textwidth]{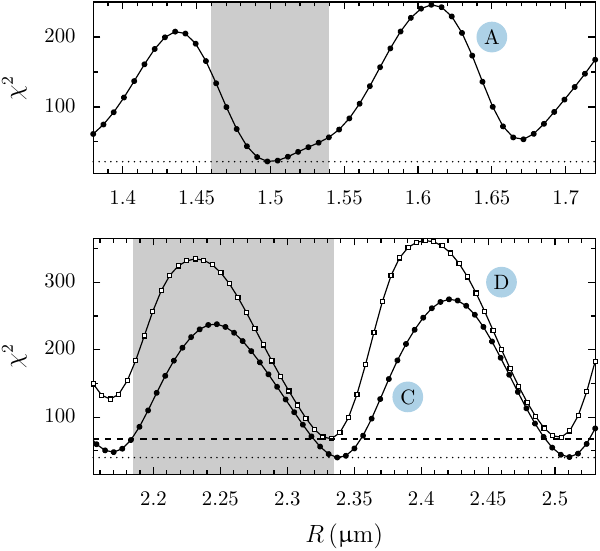}
\caption{In the bottom panel, $\chi^2$ as a function of the radius $R$ for beads C (circles) and D (squares) is shown within the MDSA theory.
The grey area depicts the error interval for the nominal radius $(2.260 \pm 0.075)\, \upmu \mathrm{m}$ provided by the bead manufacturer. The dotted (dashed) line
shows the value of $\chi^2$ for bead C (D) found from the fit and given in Tab.~\ref{tab:beadCD}.
The top panel shows, for comparison, $\chi^2$ for bead A. The grey area represents the error interval $(1.50 \pm 0.04)\, \upmu \mathrm{m}$.
}
\label{fig:chi2}
\end{figure}

Fig.~\ref{fig:chi2} displays $\chi^2$ as a function of the sphere radius for bead A (upper panel) and beads C and D (lower panel). Notice that the curves for beads C and D do not only exhibit a minimum at the optimal fit but also possess local minima at other values of the radius. The height of the dashed and dotted line serves as a reference for comparison with $\chi^2$ for the optimal fit. The grey area indicates the error interval around the nominal radius. The observed pseudo-oscillations originate in the semi-classical scattering limit from the interference between
direct reflection and radial round-trip propagation inside the sphere over 
a distance $4R$~\cite{Neto2000}, and their period is given by $\lambda_0/4n \approx 0.169\,\upmu\mathrm{m}$. 
The distance between the first two and the last two minima is found for bead C as $0.170\,\upmu\mathrm{m}$ and $0.171\,\upmu\mathrm{m}$, respectively, and for bead D as $0.164\,\upmu\mathrm{m}$ and $0.172\,\upmu\mathrm{m}$, respectively. These values are close
to the theoretically expected period considering the error bars of the fitted radii. 

The ambiguity arising from multiple minima does not have an impact on the results for bead A, as indicated in the upper panel of Fig.~\ref{fig:chi2}. Since its radius is closer to the wavelength of the trapping light, the interference effect described above is not as pronounced as in the case of the larger spheres C and D. Consequently, their fitted radii values correspond to more clearly defined global minima.

 As for beads C and D, the radii given in Table~\ref{tab:beadCD} correspond to the global minima of $\chi^2$ and they lie closer to the nominal value than radii associated with additional local minima of $\chi^2$. The minima displayed in the lower panel of Fig.~\ref{fig:chi2} at radii beyond $2.5\,\upmu\mathrm{m}$ have values of $\chi^2$ comparable to the global minima as can be seen by means of the horizontal lines. However, these radii differ from the nominal value by more than $3\sigma$ and therefore should be excluded. Even though the local minima of $\chi^2$ at a radius around $2.17\,\upmu\mathrm{m}$ clearly lie above the respective global minimum, we need to assess whether the difference between the two values of $\chi^2$ is significant. To this end, we consider the maximum of $\chi^2$ within the error interval of the fitted radii according to Table~\ref{tab:beadCD}. For bead~C, we find a maximal value of $41.2$ as compared to the value of $48.1$ at a radius of $2.169\,\upmu\mathrm{m}$. Therefore, the global minimum is significantly better than the local minimum. A similar analysis can be carried out for bead~D confirming that also in this case the global minimum is indeed significantly better.

\section{Conclusions}\label{sec:conclusions}

In conclusion, we have presented a method for determining the radii of optically 
trapped microspheres with errors on the order of a nanometer, well below 
the usual manufacturer's standard error. One of its main advantages is that it 
allows to measure a specific spherical particle \textit{in situ} during trapping experiments. 

Furthermore, we showed that micro-sized particles with $R \approx \lambda$ can be used 
to probe the optical aberrations of the experimental setup as well as to get an estimate for the height of the sphere above the coverslip, which is an experimental 
parameter hard to determine with precision.  

\section*{Acknowledgments}

We are grateful to Bruno Pontes, C\'assia S. Nascimento, Cyriaque Genet, Paula Monteiro and Suzana Frases for fruitful discussions. 
P.~A.~M.~N. and N.~B.~V. acknowledge funding from the Brazilian agencies Conselho Nacional de Desenvolvimento Cient\'{\i}fico e Tecnol\'ogico (CNPq--Brazil), Coordenaç\~ao de Aperfeiçamento de Pessoal de N\'{\i}vel Superior (CAPES--Brazil),  Instituto Nacional de Ci\^encia e Tecnologia Fluidos Complexos  (INCT-FCx), and the Research Foundations of the States of Rio de Janeiro (FAPERJ) and S\~ao Paulo (FAPESP).

\appendix

\section{Waist and Objective Filling}
\label{appendix:filling}

To ensure good trapping conditions in optical tweezers, one must guarantee that the waist of the paraxial beam is such that it illuminates the objective entrance with a significant fraction of its power distributed near the border of the lenses. In this way the marginal rays are strongly deflected, optimizing the focused beam's intensity gradient. For a Gaussian beam, this means simply overfilling the entrance. However, Laguerre-Gaussian beams require a more careful choice of waist, since most of its power is localized far from the optical axis and may leak outside the objective if the beam waist is too large.
In the experiments presented here, we have used two criteria for the choice of each mode's waist. One is the theoretical criterion presented in \cite{Fonseca2023} to control the objective filling by the ratio $r_\ell/R_{\rm obj}$, where $r_\ell$ is the radius of maximum intensity of the paraxial  $\rm LG_{0 \ell}$ beam. The second criterion was empirical, based on the limitations of the SLM. As the topological charge increases, the modulation masks necessary to reduce the beam's waist become larger, to the point that they can no longer fit inside the SLM's display. Hence, one must choose waist values that sufficiently optimize the optical tweezers' intensity gradient and, at the same time, are produced by masks that fit inside the SLM's display.

To experimentally measure the waist values of each Laguerre-Gaussian mode, we have employed a variation of the method described in \cite{viana2007}. It consists of measuring the power through a diaphragm of radius $a$ centered at the beam's axis as a function of this aperture, and then fitting the power to the integrated intensity in the diaphragm's area. In the case of a Laguerre-Gaussian mode, we have:
\begin{equation}
    P_{\ell}(R) = P_t \left[ \Gamma(|\ell| + 1) - \Gamma \left( |\ell| + 1, \frac{2 a^2}{w_0^2} \right) \right]
\end{equation}
where $P_t = (\pi/2) w_0^2 I_0$ is the total beam power in the $\ell = 0$ (Gaussian) case. Instead of an actual diaphragm, we have simulated its effect by using the spatial light modulator to divert all light outside a circle with controllable radius $R$ centered at the beam axis. The measured waist values and the ratio $r_{\ell}/R_\text{obj}$ for each mode are presented in Table ~\ref{tab:beam_waist}.

\begin{table}
\caption{Measured waist for each $\text{LG}_{0\ell}$ mode}
\centering
\setlength{\tabcolsep}{8pt} 
\renewcommand{\arraystretch}{1.2}
\begin{tabular}{ccc}
\hline\hline
 $\ell$ & $w_0 \, (\mathrm{mm})$ & $r_{\ell}/R_{\rm obj}$ \\
 \hline
 0  &$2.150 \pm 0.004$& -\\
 $\pm 1$&$1.732 \pm 0.008$& $0.44$\\
 $\pm 2$&$1.204 \pm 0.004$& $0.43$\\
 $\pm 3$&$1.171 \pm 0.005$& $0.51$\\
 $\pm 4$&$1.039 \pm 0.007$& $0.52$\\
 $\pm 5$&$0.948 \pm 0.003$& $0.54$\\
 $\pm 6$&$0.867 \pm 0.002$& $0.54$\\
 $\pm 7$&$0.806 \pm 0.002$& $0.54$\\
 \hline\hline
\end{tabular}
\label{tab:beam_waist}
\end{table}

\section{Multipole expansion of the optical force}\label{sec:force_components}

Here, we present the series expansion of the optical force components
for a Laguerre-Gaussian beam within the MDSA+ theory \cite{dutra2014}. 

We use a dimensionless force $\mathbf{Q} = \mathbf{F}/(n_\text{w} P/c)$, 
with the laser beam power $P$, the speed of light $c$ and the refractive index in the sample region $n_\text{w}$.
The force is given by subtracting the loss  at the sphere (-$\mathbf{Q}_s$) from the total extinction ($\mathbf{Q}_e$) 
\begin{equation}
\mathbf{Q}(\mathbf{R}) = \mathbf{Q}_s + \mathbf{Q}_e
\end{equation}
and it is calculated at the position $\mathbf{R} =  \mathbf{R}(\rho, \phi, z)$ of the spherical particle with respect to the focal spot. 

First, we present the components of the scattering force.
The axial part of the scattering force is given by
\begin{widetext}
\begin{equation}
\begin{aligned}
Q_{sz}^{(\sigma, \ell)} = -\frac{8\gamma^2}{A_\ell N_s} \mathrm{Re} 
\sum\limits_{j=1}^\infty \sum_{m=-j}^j 
&
\left[ \frac{\sqrt{j(j+2)(j-m+1)(j+m+1)}}{j+1} 
(a_j a_{j+1}^{*} + b_j b_{j+1}^{*}) G_{j,m}^{(\sigma, \ell)} G_{j+1,m}^{(\sigma, \ell)*} \right. 
\\
& \quad \left.
+ \sigma m \frac{2j+1}{j(j+1)} a_j b_j^{*} \left|G_{j,m}^{(\sigma, \ell)}\right|^2 \right]
\label{eq:Qsz}
\end{aligned}
\end{equation}
where $a_j$ and $b_j$ are the Mie coefficients and $A_\ell$ defines the so-called filling factor given by
\begin{equation}
A_\ell = 8(2\gamma^2)^{|\ell|+1}
\int_0^{\sin\theta_\text{m}}  t^{2|\ell| +1} e^{-2\gamma^2 t^2}
\frac{\sqrt{(1-t^2)(N_\text{s}^2 - t^2)}}
	{\left(\sqrt{1-t^2} + \sqrt{N_\text{s}^2 - t^2}\right)^2}\,
	\mathrm{d}t \,.
\end{equation}
The transverse components of the scattering force are given by
\begin{equation}
\begin{aligned}
\left\{\begin{array}{c}
Q_{s\rho}^{(\sigma, \ell)} \\
Q_{s\phi}^{(\sigma, \ell)}
\end{array}\right\}
 = \frac{4\gamma^2}{A_\ell N_s} 
\left\{ \begin{array}{c}
 \mathrm{Im} \\
 - \mathrm{Re}
 \end{array}\right\}
 \sum\limits_{j=1}^\infty \sum_{m=-j}^j 
 & \left[\frac{\sqrt{j(j+2)(j+m+1)(j+m+2)}}{j+1} 
(a_j a_{j+1}^{*} + b_j b_{j+1}^{*}) \right.
\\
&\hspace{5em} \times
\left( G_{j,m}^{(\sigma, \ell)} G_{j+1,m+1}^{(\sigma, \ell)*}
 \pm G_{j,-m}^{(\sigma, \ell)} G_{j+1,-m-1}^{(\sigma, \ell)*}\right)
\\
&  \quad - 
\left. 2\sigma\frac{2j+1}{j(j+1)} \sqrt{(j-m)(j+m+1)} \mathrm{Re}(a_j b_j^{*})
 G_{j,m}^{(\sigma, \ell)} G_{j,m+1}^{(\sigma, \ell)*}
 \right]
\end{aligned}
\label{eq:Qsrho}
\end{equation}
where the upper sign corresponds to the radial force component, while the lower sign is 
for the azimuthal part. 
The multipole coefficients of the circularly polarized Laguerre-Gaussian 
beam, including optical aberrations, are given by
\begin{equation}
\begin{aligned}
G_{j, m}^{(\sigma, \ell)}(\mathbf{R}) = (\sqrt{2}\gamma)^{|\ell|} 
\int_0^{\theta_m} \mathrm{d}\theta  
\sqrt{\cos(\theta)} \sin^{|\ell|+1}(\theta)
e^{-\gamma^2 \sin^2(\theta)}
d_{m, \sigma}^{j} (\theta_w) 
 f^{(\sigma, \ell)}_m (\mathbf{R})
e^{ik_w\cos(\theta) z + i\Psi_\text{g-w}} \,.
\end{aligned}
\label{eq:G}
\end{equation}

\end{widetext}

The coefficient $f_m^{(\sigma, \ell)}$ accounts for 
the astigmatism. The explicit expression for a Gaussian beam can be found in 
Eq.~(8) of \cite{dutra2014} and for a Laguerre-Gaussian beam we obtained 
\begin{equation}
\begin{aligned}
f^{(\sigma, \ell)}_m (\mathbf{R}) = &
\sum_{s=-\infty}^\infty (-i)^{s} J_s\left(2\pi A_\text{ast} \frac{\sin^2\theta}{\sin^2\theta_0}\right)
 \\
&\times 
J_{2s +m -\sigma-\ell}(k \rho \sin\theta) e^{2is(\phi_\text{ast} - \phi)}\,.
\end{aligned}
\end{equation}
In the absence of astigmatism $A_\text{ast} = 0$, only the $s=0$ term
contributes to the sum and the coefficient reduces to 
\begin{equation}
f^{(\sigma, \ell)}_m (\mathbf{R}) = J_{m -\sigma-\ell}(k \rho \sin\theta)\,.
\end{equation}

Next, we present the expressions for the extinction force. 
The axial part is given by 
\begin{equation}
Q_{ez}^{(\sigma, \ell)} = \frac{ 4\gamma^2}{A_\ell N_s} \mathrm{Re} \sum\limits_{j,m} (2j+1)(a_j + b_j) 
G_{j,m}^{(\sigma, \ell)} \left(G^{(\sigma, \ell)'}_{j,m}\right)^{*}. 
\label{eq:Qez}
\end{equation}

The coefficient $G^{(\sigma, \ell)'}_{j,m}$ can be expressed in terms of the
multipole coefficients \eqref{eq:G} by applying the recursion relation for Wigner-$d$ matrix elements
\cite[p.~90]{varshalovich1988}

\begin{widetext}

\begin{equation}
G^{(\sigma, \ell)'}_{j,m} = 
\frac{\sqrt{j(j+2)\left[(j+1)^2 -m^2 \right]}}{(2j+1)(j+1)}
G_{j+1, m}^{(\sigma, \ell)} 
+
\frac{\sqrt{(j^2 -m^2)(j^2-1)}}{j(2j+1)}  G_{j-1, m}^{(\sigma, \ell)} +
\sigma \frac{m}{j(j+1)} G_{l,m}^{(\sigma, \ell)}\,. 
\label{eq:G_prime}
\end{equation}

The transverse components of $\mathbf{Q}_\mathrm{e}$ are given by
\begin{equation}
\left\{\begin{array}{c}
Q_{e\rho}^{(\sigma, \ell)} \\
Q_{e\phi}^{(\sigma, \ell)}
\end{array}\right\} 
= \frac{2\gamma^2}{A_\ell N_s} 
\left\{\begin{array}{c}
\mathrm{Im} \\
- \mathrm{Re}
\end{array}\right\} 
 \sum\limits_{j,m} (2j+1)(a_j + b_j) G_{j,m}^{(\sigma, \ell)}
  \left(G^{-, (\sigma, \ell)}_{j,m+1} \mp G^{+, (\sigma, \ell)}_{j,m-1}\right)^{*} \,.
\label{eq:Qerho}
\end{equation}
The negative (positive) sign corresponds to the radial (azimuthal) component. The multipole 
coefficients $G^{\pm, (\sigma, \ell)}_{j,m-1}$ can also be expressed in terms of the coefficients
$G^{(\sigma, \ell)}_{j,m-1}$ by again applying recursion relations
\begin{equation}
\begin{aligned}
G^{\pm, (\sigma, \ell)} _{j,m} &= 
\mp \frac{\sqrt{(j\pm m)(j \pm m+1)(j^2 -1)}}{j(2j+1)} G_{j-1,m}^{(\sigma, \ell)}  + 
\sigma \frac{\sqrt{(j\mp m)(j \pm m+1)}}{j(j+1)} G_{j,m}^{(\sigma, \ell)}  \\
& \qquad \pm
\frac{\sqrt{(j\mp m)(j\mp m+1)((j+1)^2 -1)}}{(j+1)(2j+1)} G_{j+1,m}^{(\sigma, \ell)} \,.
\end{aligned}
\label{eq:G_pm}
\end{equation}
Making use of the recursion relations \protect\eqref{eq:G_prime}
and \eqref{eq:G_pm} reduces drastically the need to explicitly evaluate integrals of the form \eqref{eq:G}
and thus significantly improves the runtime of the numerical calculations.
\end{widetext}

\bibliography{lgbeam}

\end{document}